\documentclass[aps,pre,preprint,superscriptaddress,longbibliography]{revtex4-1}

\usepackage{amsmath,amssymb,pifont}
\usepackage{graphicx,color,subfigure,bm}

\begin{document}

\title{Pattern formation by droplet evaporation and imbibition in watercolor paintings}


\author{Jorge Gonz\'alez-Guti\'errez}
\affiliation{Departamento de Termofluidos, Facultad de Ingenier\'ia, Universidad Nacional Aut\'onoma de M\'exico, Av. Universidad 3000, Coyoacan CDMX 04510, M\'exico}

\author{Octavio Moctezuma}
\affiliation{ecretaria de Cultura Federal, Ciudad de M\'exico,  M\'exico.}

\author{Ver\'onica Angeles}
\affiliation{Instituto de Investigaciones en Materiales, Universidad Nacional Aut\'onoma de M\'exico, Apdo. Postal 70-360, Coyoacan, Ciudad de Mexico 04510, Mexico}

\author{Maricarmen Rios-Ramirez}
\affiliation{CINVESTAV-Monterrey, Apodaca, 66600, M\'exico}

\author{Sandra Zetina}
\affiliation{Instituto de Investigaciones Est\'eticas, Universidad Nacional Aut\'onoma de M\'exico, Coyoacan CDMX 04510, M\'exico}

\author{Alvaro Marin}
\affiliation{Physics of Fluids Group, University of Twente, Enschede, The Netherlands}

\author{Roberto Zenit}
\email{zenit@unam.mx; zenit@brown.edu}
\affiliation{Instituto de Investigaciones en Materiales, Universidad Nacional Aut\'onoma de M\'exico, Apdo. Postal 70-360, Coyoacan, Ciudad de Mexico 04510, Mexico}
\affiliation{School of Engineering, Brown University, 184 Hope St, Providence, RI 02912, USA}

\date{\today}

\begin{abstract}
Watercolor or aquarelle is one of the oldest painting methods in which pigments suspended in an aqueous liquid are deposited over a substrate, typically an absorbing material such as paper. Artists manipulate the inks and paper to create a wide variety of textures which serve as the building pieces of the painting composition. The physical processes which lead to pattern formation have not been studied in any depth, in spite of being closely related to flows in other contexts. Identifying and understanding these processes is of imperative importance to restore and preserve watercolor paintings. Here, we report an experimental study of the effect of the pigment concentration and paper humidity on the pattern formation derived from evaporation of droplets of watercolor paintings.  Optical analysis reveals the formation of color gradients, stratifications, flat regions, borders, dendritic shapes, and radial tips. We found that droplet evaporation on dry paper forms ring-shaped stains which resemble to the \emph{coffee-stain effect} regardless of the nature of the pigment.  The mean pixel intensity of such deposits follows an exponential function that saturates at high concentration, while the thickness of the coffee ring increase for watercolor inks containing colloidal
 particles and does not change for non-colloidal .  Our experiments reveal that water distribution on the paper surface,  and not the volumetric absorption of water on the paper,  determine the structural characteristics   of watercolor stains.  We show evidence that the cornerstone in the creation of complex patterns in watercolor paintings is driven by the coffee ring effect and imbibition processes. Our findings aim to serve as framework for further investigations of the complex processes involved in this ancient art form and could guide restoration processes needed to preserve the heritage value of historical watercolor artworks.
\end{abstract}

\maketitle



\section*{Introduction}

The restoration of deteriorated paintings safeguards the cultural, historical, and the commercial value of a piece of art. Sophisticated techniques such as fourier transform infrared microscopic \cite{Van}, fluorescence spectrometry \cite{Klockenkamper}, Raman microscopy \cite{Burgio}, among many others, give us information about the physicochemical state of a piece. However, the physical processes that lead to the formation of patterns and textures in watercolor paintings have not been investigated in detail to date. In particular, further understanding of the mechanics of watercolors would be of great value in choosing the best strategies to preserve the technical merit of masterpieces.

Watercolor painting is perhaps one of the oldest human art forms \cite{Brett}. The cave paintings of paleolithic Europe were painted using water-based colors; many manuscript illustrations in ancient Egypt and in the European Middle Ages were also painted with what we now call watercolour. It is also the technique taught to most children around the world as a introduction to painting and art. Though historically the use of watercolor has evolved always with different artistic schools, an illustration of some notable pieces is shown in Fig. 1. In all of them, artists express their perceptions, sensations and emotions  of reality, which contemplates the place where humanity exists and the alternative reality that is created in the mind.

The technique consists of applying a paint, pigments and a binder suspended in water, to a surface, usually paper \cite{Mayer}.  In most cases, the pigments are not soluble in water, so the paint is in fact an aqueous suspension. The binder is commonly gum arabic (a complex mixture of glycoproteins and polysaccharides); it helps the pigments remain suspended in water and it binds pigments to the paper after drying \cite{Abdulsalam}. Prewetting of the painting support (paper)  is often done, carried out by immersing the paper into a water container,  using a  moisturized brush, or depositing water directly. In some cases, alcohol is added to change the evaporation rate and the contact properties of the inks. Professional artists learn, by repetition and lengthy experimentation, how to manipulate the paint and surfaces to create textures needed for their compositions.

The magic happens  not only when the artist applies the paint but also while drying of the paint occurs. Depending on the environmental humidity and the water content in the paper, the dye might attach to the paper or pass through it. Once the solvent evaporates, different textures emerge. However, as occurs in others forms of art, there is no general rule for the creation of watercolor pieces. Instead, the watercolor technique is so rich that allows the inherent creativity of each artist to manifest in a way dictated by the choice the different parameters involved in the process (ink and paper type, application techniques, humidity, timings, etc.). If one looks carefully at watercolor paintings it can be perceived that, regardless of the artistic composition and the historical context,  all of them show similar characteristic textures and patterns, some of which are emphasized in Fig. 1. In particular, we can identify well-defined edges, gradients of color, flat surfaces with uniform deposition of pigments, small saturated radial tips, large radial tips, and dendrites.

Knowledge of how water distributes in paper is of fundamental importance for the paper making industry and operations such as coating, surface sizing, packaging, and ink-jet printing. Therefore, there is substantial interest to understand the physical processes involved in the wetting and drying of paper. These seemingly simple mechanisms include phenomena such as: Capillary transport and vapor diffusion through complex porous media, droplet wetting of deformable and porous substrates, water absorption to fiber-like structures, etc. \cite{Rounsley,Liang,Nilsson1,Nilsson2,Hashemi, Radhakrishnan, Massoquete}. The large of number of phenomena involved in the process makes it impossible to apply theoretical models. Numerical simulations might eventually help to understand how all these processes are intertwined, but up to now only oversimplified cases can be solved \cite{Harting2016}. From the experimental side, techniques such as nuclear magnetic resonance (NMR) imaging\cite{Nilsson3,Li,Harding}, solute exclusion \cite{Scallan}, and thermoporosimetry\cite{Maloney} allow to explore mass transport phenomena in paper and cotton fibres.

The process of water evaporation  on a dry surface is dictated by complex mass transport mechanisms. The ``simplest'' scenario consists of a droplet pinned on a flat surface. A competition between capillary flow (driven by capillary forces) and Marangoni flows (driven by surface tension gradients) emerge during the droplet evaporation \cite{deegan2000contact,hu2006marangoni}. Both capillary and Marangoni flows result from the diffusive evaporation process. The capillary flow results from a combination of evaporation and contact line pinning. In order to retain the pinning condition, while maintaining a minimum surface, the liquid is transported radially outwards. If colloidal particles dispersed in the liquid phase, a ring-shaped stain (\emph{coffee ring}) appears: a ring-shaped structure at the edge of the droplet \cite{deegan1997capillary,Marin2011:order}. Marangoni flows, on the other hand, appear due to the appearance of surface tension gradients at the interface which can be created by surfactants \cite{still2012surfactant,sempels2013auto,Marin2016:surfactant}, by disolved solutes \cite{Marin2019:salt} and temperature gradients \cite{hu2006marangoni,hu2005marangoni}.



Solely hydrodynamics can explain canonical cases like the classical coffee-stain case as long as particles behave as hard spheres and Marangoni flows are negligible, i.e. all streamlines end at the contact line. More generaly, nonetheless, the morphology of the final deposit depends on complex  physico-chemical processes mainly involving particle-wall, particle-interface and particle-particle interactions that can lead to a large variety of complex aggregates such as amorphous peripheral ring, dendritic shapes, rosettes, scallops, Chinese arrows, zigzag patterns, undulated branches, and interlocked chains \cite{gorr2014pattern,chen2010complex,gorr2012characteristic,yakhno2011complex,carreon2017patterns,pauchard1999influence}.  Moreover, the complex structural characteristics of the deposit patterns have been exploited to the diagnostic  health problems \cite{Sefiane}, to explore cell motility \cite{Nellimoottil,Rios}, calorimetric properties in membranes \cite{gonzalez2017calorimetric},  authenticate consumable beverages \cite{gonzalez2017technique}, among many others uses.

This work aims to study the most predominant phenomena that occur during the production of watercolor painting with the intention to serve as a framework to more detailed studies and a guide for practical applications as restoration, ink-jet printing or even new artistic directions. More precisely, we report an experimental study of the effect of the pigment concentration and paper moisture on the pattern formation derived from the droplet evaporation in watercolour paintings. Depending on the pigment concentration and the water distribution on the paper, many complex structures such as well-defined edges, gradients of color, flat surfaces with uniform deposition of pigments, small saturated radial tips, large radial tips, and dendrites emerge after the evaporation of the droplets. Interestingly, all these patterns nest in the art pieces shown in Fig. 1. We found that the coffee rings effect is a universal effect that appear during the patter formation regardless of the nature of the pigment, the concentration or water density in the paper.  We found that the mean pixel intensity of the coffee ring stains created on dry paper follows an exponential function that saturates at high concentration, while the corresponding thickness increase for smaller  particles and does not change for larger ones.  The experiments show that the formation of patterns on wet paper is driven by three different stages of evaporation of the paper: the formation of water bodies (stage I), the creation of stagnant water (stage II), and water in bulk (stage III).  Interestingly, the proportion of water in paper is not a relevant parameter for the pattern creation.  We believe that our findings could help to improve the restoration processes to preserve the heritage value of watercolor paintings as well as serve as framework for further studies on the drying process of stains over wet porous surfaces, which are so commonly seen in a large variety of applications.

\section*{Results}

\subsection*{The effect of the pigment concentration on dry paper}

A sequence of images of the pattern formation from the evaporation of droplets is shown in Fig. 2.  This intricate process begins when the drop is deposited on the surface of the paper.  The droplet spreads until a mechanical equilibrium is reached and the contact line remains stable with a well-defined apparent contact angle (see Fig 2a). During the evaporation process, pigments are deposited on the paper surface  forming, in most of the cases, well-defined ring-shaped stains (see Fig. 2b).

The patterns left by droplets evaporation at different pigment concentrations $\phi_p$ are shown in Fig. 3a.  Except for only two structures, all of them are composed by a ring-shaped stain surrounded by a quasi-uniform deposition of pigments. The radial density profile $I(r)$ reveals these structural characteristics. For example, Fig. 3b shows the corresponding radial density profile of a stain containing pink pigment at $\phi_p$=0.5   (where $\phi_p$ is the rate between the diluted solutions and the stock solution which correspond to the saturation concentration of Mars Black in water).  A zoom of the inner deposits and the coffee ring stains are shown in Fig. 3c.

We analysed the mean pixel intensity $\bar{\mu}$ of the inner regions (the area that is surrounded by the coffee ring  in order to quantify the morphological changes produced by varying the pigment concentration. The $\bar{\mu}$  of deposits formed at a different concentration on dry paper are plotted in Fig. 3d.  Interestingly, we found that $\bar{\mu}$ change exponentially as follows:
\begin{equation}
\bar{\mu}=  -\bar{\mu}_0 e^{(- \phi_p / k )}+\bar{\mu}_s,
\end{equation}
where $\bar{\mu}_0$ is the basal value of the paper,  $\bar{\mu}_s$ is a value where the mean pixel intensity saturates,  and $k$  is the characteristic concentration $\phi_c$ at which the magnitude $\bar{\mu}_0$ diminishes by a $1/e$ factor. The normalized mean pixel intensity  $\gamma$ show that, regardless of the nature of the pigments, the pixel intensity universally follows an exponential function that saturates at a high pigment concentration (see the inset of Fig. 3d). This means that regardless of the pigment color, the deposition process follows the same dynamics. The results are in agreement with previous reported works on the saturation of aggregates in deposits formed by protein solutions \cite{gonzalezproteins}.

Next, we analyze the coffee ring thickness as a function of the pigment concentration.  One would expect that the ring thickness should increase as the particle concentration increases. That is precisely what we find for pigments with small particles (within the colloidal range). However, the pigment concentration does not change significantly for inks containing larger  non-colloidal particles (black and blue pigments, see Fig. 3f).  This behavior has a clear correlation with the size and composition of the particle pigments. Most particles of the black and blue pigments have significantly larger size, slightly above the colloidal range, and significantly larger densities as the rose pigments. Consequently, by computing the settling velocities (see Fig. 1s in the  supplementary information) we can conclude that all particles in red and pink pigments remain suspended and contribute to the formation of the coffee ring. While most of the particles in the black and blue pigments sediment in the order of seconds, there is a significant amount of particles -- specially in the black pigment-- that will certainly reach and contribute to the ring. The reason why its thickness is insensitive to the increase of that fraction of black particles (43.3\% of particles of 258 nm, see table 1) is not completely clear to us. One reason could be a very efficient three-dimensional packing of the particles, which might give rise to an growth of the ring in height but not in the radial direction. Another potential reason could be a preferential absorption of the smaller particles in the solution into the porous matrix.

\subsection*{The effect of moisture on the pattern formation}

Before analyzing the formation of watercolor patterns on wet paper, we must remark that the same content of water in the paper can produce three different types of water distribution: water bodies (I), stagnant water (II) and water in bulk between pores and fibers (III). The water bodies (I) are amorphous water puddles (asymmetrical water bodies) that arise over the paper immediately after the wetting process. Interestingly, they evolve to the radial symmetry during the evaporation to form symmetrical water bodies. For the stagnant water case (II) the liquid is within the superficial paper cavities (formed by the roughness of the paper) that produces a high gloss appearance. The water in bulk (III), on the other hand, is water placed between the pores and fibers which produces a dull appearance on the paper. A complete analysis of how these types of distribution of water appear during three stages of the evaporation in the watercolor paper is found in the Methods section.

Fig. 4a shows a sequence of images of radial pattern formation from the deposition of a drop of watercolor on symmetrical water body (Type I). At first, the natural imbibition process, which arises from the deposition of the ink drop, allows the ink to rapidly diffuse radially into the water body  ($t/t_c=0.06$, where $t_c$ is the characteristic time at which the mass of  a droplet on a dry paper diminishes by a $1/e$ factor at T= 22 $^\circ$C and relative humidity of 35\%, see Fig. 2s in the supplementary information. For a droplet of 50 $\mu l$,  $t_c$=2490s). Afterwards, a second slow diffusion process emerges, from the center of the water body towards the periphery ($t/t_c=0.12-0.361$). This process results from capillary fluxes, which are induced as a consequence of the higher rate of evaporation in the water mass occurring on its periphery which feeds the diffusion of ink over the stagnant water located around the water body. We must remark that similar pattern formation emerges from ink deposition on asymmetrical water body (Type I) which emerge in the early stage of drying, see Fig. 3s in the supplementary information.

The morphological changes that the pattern undergoes during its formation can be quantified by the configurational entropy which is a measure of the heterogeneity of the intensity distribution along the image, where higher values correspond to higher complexity, see Method Section. Fig. 4b. shows that the entropy increases during the initial rapid imbibition process. Interestingly, the maximum complexity of the structure is reached once the structure reaches its maximum diameter (see blue line in Fig. 4b). Then, during the slow diffusion process, the complexity of the structure decreases. This is due to the fact that the slow diffusion of ink drags the pigments homogeneously from the center towards the periphery, smoothing the radial tips and reducing the heterogeneity of the structure.  

In order to explore the role of the loss of water on the pattern formation during the drying process, Fig. 4c shows the evolution of the relative mass $m/m_0$ (where $m_0$ corresponds to the initial mass of the dry paper) as a function of time. The abrupt change in the upper curve corresponds to the moment when the drop of ink is deposited on the paper ($t/t_c=0$), which is more evident in the abrupt change in its derivative form (see the blue line in Fig. 4c). Note that the mass of water decays monotonically in time throughout the process of watercolor ink pattern formation. This is because the two mass transport mechanisms (imbibition diffusion and slow diffusion) are not governed by the evaporation process.  

The second type of water distribution, stagnant water in stage (II), is shown in Fig. 4d, which shows the process of formation of a pattern. At the beginning, the drop of watercolour ink is deposited on the paper forming a body of quasi-symmetric water ($t/t_c=0$). The imbibition process between the ink and the stagnant water allows the growth of large stain with irregular borders characterized by sharp tips ($t/t_c=0.016$). However, the tips overlap and deform during their growth ($t/t_c=0.0301$).  Afterwards, the slow diffusion process emerges from the center of the water body towards the periphery, generating a ring around the structure($t/t_c=1.35$). Finally, the watercolor ink body vanishes ($t/t_c=1.8$).  Interestingly, for early stage of evaporation of stagnant water the  ring around the structure is is narrower, as you can see in Fig. 4s in the supplementary information.

Fig. 4e shows the evolution of the entropy during  the pattern formation corresponding to Fig. 4d.  Interestingly, although the structure reaches its maximum diameter quickly (see blue line in Fig. 4e), the heterogeneity continues to grow due to the formation of the ring. Afterward, the entropy is reduced because the water of the ink body inside the structure disappears. The evolution of the relative mass $m/m_0$ as a function of time is shown in Fig. 4f. The water content changes during the whole process of watercolour ink pattern formation. The discontinuity in the curve corresponds to the moment when the drop of ink is deposited on the paper.  The absence of changes in the derivative form of the relative mass shows that the two mass transport mechanisms (imbibition diffusion and slow diffusion) that emerge during the pattern formation are not governed by the evaporation process (see blue line in Fig. 4f).

In the third type of water distribution (III) the capillary flow feed the formation of dendrites around a coffee ring. Fig. 4g shows a sequence of images of the formation of a pattern from the deposition of a drop of watercolour on paper in this last drying stage. The watercolour ink drop is attached on the wet paper, forming a contact angle much lower than that observed on dry paper. The lower contact angle might explain why the capillary flow seems stronger in this case, leading to a fast formation of a coffee ring.  The watercolour ink diffuses slowly from the center of the drop towards the periphery. Then, by capillarity, a process of slow diffusion that forms dendrites emerges ($t/t_c=0.15$). Finally, once the drying process has finished, the whole structure appears ($t/t_c=3.01$).

The entropy quantify the dendritic growth of the structure and the formation of the coffee ring. Fig. 4h. shows that the heterogeneity  increases during dendritic growth. While the diameter of the structure ceases to grow (see blue line in Fig. 4h), its complexity continues increasing. This is due to the slow saturation of pigment near the contact line of the drop. Afterwards, the  entropy decreases because the droplet fades away. However, the complexity increases again due to the emergence of the coffee ring. Fig. 4i shows the corresponding evolution of the  relative mass $m/m_0$  as a function of time during the formation of stains. The derivative of surface water density as a function of time is shown by the blue curve in Fig. 4i. As in the previous case, the abrupt change in both curves correspond to the instant the ink drop is deposited on the paper. Note that both quantities change continuously  during the whole process of pattern formation. The presence of the wet paper would typically partially or totally neutralize the evaporative singularity typically present at contact lines \cite{deegan1997capillary}. However, it is well-known that such a singularity at the contact line is not necessary for a coffee ring to occur \cite{deegan2000contact,Gelderblom2012}, and the ring-shaped patterns observed in (II) (Fig. 2d) and (III) (Fig. 2g) are good examples of such an effect.

In Fig. 5a we show the growth process of structures containing non-colloidal pigments in the three types of water distribution. Fig. 5b shows the evolution of the configurational entropy and relative mass $m/m_0$ as a function of time during structure formation.  It should be noted that, comparing the formation of patterns with colloidal and non-colloidal pigments, it is possible to assume a lower efficiency of the mass transport mechanism in the latter. In fact, this deficiency is evident when comparing the thickness of the coffee rings formed during the drying of ink droplets on dry paper. 

The structural characteristics of watercolor patterns depend on the water distribution of the paper on which the ink is deposited, and not on the content of water in paper surface. In fact,  the three water distributions on paper can form slightly different patterns if the ink deposition is carried out in the different stages of water distributions drying. Fig. 6a shows patterns formed from ink deposition for the early and late stage of drying of the three water distributions. Regardless of the diameter of the pigments, the first type of water distribution (I) allows the formation of patterns with radially-oriented sharp tips. Patterns without radial symmetry are generated by the ink deposition on amorphous water bodies that appear in the early stage of drying of the first type of water distribution (early-stage). In contrast, patterns with radial symmetry are formed from ink deposition on symmetrical water bodies (late-stage).   These stains are the bigger ones of the set of patterns formed from the evaporation of watercolor ink in wet paper, see Fig. 6b-c.

From the second type of water distribution (II) the colloidal and larger diameter pigment patterns are clearly distinguishable. Inks with colloidal pigments form a pattern composed of a uniform deposit surrounded by a clear ring. The coffee-ring is the most evident difference between stains formed in the early and late stage of drying of the second type of water distribution. In contrast, watercolor inks with non-colloidal pigments generate the same patterns. These are composed by a small uniform deposit of high density surrounded by a large region of low pigment density.  The third type of water distribution (III)  allows the formation of dendrites around a well-defined coffee ring from inks containing colloidal particles. Interestingly, inks containing non-colloidal pigments only form uniform, high-density deposits.  Fig. 6b-c  shows that the average diameter of structures decreases for the last two last types of water distribution.

\section*{Discussion}
The restoration of art involves the deep knowledge of materials and techniques in order to preserve an artistic piece in its original state. In general, a painting is composed by the \emph{imprimatura}, the bottom of the bolus, the binder, the pigments and in some cases, the protective layer of the works. The analysis of such materials allows generating an efficient strategy to carry out a restoration process. Invasive restoration has been a technique commonly used to preserve and recover masterpieces. This technique consists of reconstructing a complete work through the use of washable inks that allow reverting unwanted mistakes during the rehabilitation of a painting. Typically, invasive restoration begins with adhering a similar material to the original ones. The adhesion is done with wax or polymers that can be removed without damaging the original work. Thereafter, the restorer must consolidate the work through painting techniques based on inks that are capable of reproducing the morphological characteristics that nest in the works. In the literature, one finds many successful cases about the use of this resource for acrylic paintings.

The restoration based on invasive techniques is not completely appropriate for watercolor painting because these works are created, in general, with washable paints. Indeed, we found many easily identifiable problems with the restoration based only on the application of inks. For example (1)  the appearance of delineations is practically universal on dry paper due to the coffee ring effect. Accordingly, it is impossible to obtain a completely uniform coating (border free). On the other hand, (2) high  pigments concentrations could generate an unwanted roughness on the paper.  In the restoration based on the wetting of the cotton paper,  (3) the wetting also could dissolve pigments and irreversibly modify regions that do not need intervention. (4) The imbibition process could generate unwanted tones due to a mixing process between pigments fixed on the paper and pigments suspended in the aqueous solution.  (5) The dendritic shapes, that grow randomly on paper could invade some layers of pigments. Finally, although the type of water distribution determines the structural characteristics of the spots, the ``early''  and ``late''  states of evaporation determine the diameter of the structures, as we can see in Fig. 6. Unfortunately, this results in a poor  control of the diameter of the structures.

The understanding of pattern formation from the evaporation watercolor inks has many inherent advantages to carry out a restoration process. (1) Consider the color intensity profile for each pigment could facilitate the correct and systematic reproduction of colors in watercolors masterpieces.  (2) The use non-colloidal particles is recommendable to avoid changes in the thickness of the coffee ring, while (3) the use of  pigments with particles in the colloidal range favors the increase in the thickness of the coffee ring. (4) The distribution of water on the paper surface evolves in three stages: formation of water bodies, stagnant water and water in bulk. (5) The formation of watercolour patterns depends on the evaporation stage in which the watercolour ink is deposited, and not on the surface density of the water. (6) The water bodies that appear in the first evaporation stage (I) favour the formation of radial patterns. (7) From the second evaporation stage onwards, the patterns created with Brownian pigments and those with larger diameters are clearly distinguishable. (8)  Colour gradients arise from the sedimentation of non-colloidal particles. (9) The comparison of the pattern formation processes of ink containing colloidal vs non-colloidal pigments, confirms the fundamental role of capillary flows. There is a lower efficiency of the mass transport for larger-sized pigments. (10) The creation of dendrites emerges in the last stage of evaporation. 

In conclusion, we study the relations between many patterns in watercolor painting (color gradients, stratifications, flat regions, borders, dendritic shapes, and radially-oriented sharp tips) and three control parameters: pigment type, the pigment concentration and water distribution of the paper. We found that the mean pixel intensity of the inner regions of stains formed on dry paper exponentially saturates with the concentration. Interestingly, a simple parameter such as the configurational entropy can describe fully the morphological changes during the pattern formation. We show that the coffee ring effect and the imbibition process are universal mechanisms behind the pattern formation in watercolour paintings. We claim that these processes are, in fact , co-authors of the watercolor paintings.  Overall, the knowledge of the effect of these parameters could avoid adverse physical effects capable to generate permanent and irreversible damage during the restoration of watercolor works.

\section*{Methods}

\subsection*{Watercolor ink preparation}
Four watercolor pigment (Mars Black, Ultramarine Blue, Rose Genuine, and Permanent Rose  of Winsor and Newton) were used to prepare ink solutions. These powders  were dissolved in deionized water (Mili-Q, 18.2 $M \Omega \cdot cm$) at a concentration of 0.0075 g/mL at 22 $^\circ$C. This concentration is very close to the saturation concentration of the Mars Black pigments. The stock solutions were diluted according to the desired concentrations: 0.00375, 0.001875, 0.000937, 0.000468, 0.000234, and 0.000117 g/mL.  The solutions were stored at 2 $^\circ$C, and thermalized to room temperature before deposition. We analyze the formation of deposits at a different relative concentration of proteins $\phi$=$B/A$, where A and B stands for the stock solution and the diluted solutions, respectively.

\subsection*{Measurements of the size of pigments}
The particle size distribution of the pigments in the watercolor inks were measured by dynamic light scattering using a Zetasizer Nano ZSP (Malvern Instruments Ltd). Table 1 shows the technical data of the pigments.

\begin{table}[htbp]
\begin{tabular}{lllll}\hline \hline  
Pigment        & Element                                            & Particle size  (nm) & Error & (\%)  \\
\hline  
Mars Black     & Synthetec iron oxides                              & 1528                & 504 & 53.4 \\
               &                                                    & 258                 & 71  & 43.3 \\
               &                                                    & 5322                & 374 & 3.3  \\
Ultramarine    & Sulfur-containing sodium aluminum silicate & 3810                & 78  & 55.4 \\
               &                                                    & 1669                & 401 & 44.6 \\
Permanent Rose & Quinacridone                                       & 272                 & 73  & 83.8 \\
               &                                                    & 99                  & 19  & 16.2 \\
Rose Genuine   & Rubia tinctorum                                    & 164.5               & 29  & 50.4 \\
               &                                                    & 573                 & 118 & 49.6\\
\hline
\hline
\end{tabular}
\caption{Watercolor pigments. (\%) and Error indicates percentage of pigments in the ink and the std, respectively. }
\label{tab:my-table}
\end{table}

\subsection*{Drop evaporation}
The droplets of  inks were placed onto dry and wet watercolor paper slides (75x75mm) using a micropipette. We used cold press paper made out of  25\% cotton at 300g/$m^2$ from Watercolour-FABRIANO. The volume of the drops was 50 $\mu l$ in all cases. The droplets were evaporated under controlled ambient conditions: T= 22 $^\circ$C and relative humidity of 35\%. The time for complete evaporation of a droplet $t_c$ on dry paper was around 2500s.  The evaporation process was recorded  at 30 fps with a digital camera (Nikon Digital, SLR Camera D3200). The deposits were observed after evaporation in ambient conditions using a microscope (Velab, VE-M4, 4x and 10x).

\subsection*{Image analysis}
We use the radial density profile $I(r)$ to carry out the structural analysis of the patterns. This quantity describes a profile of integrated intensities produced by concentric circles as a function of radial distance.  For 2D objects this quantity is given by the following expression:
\begin{equation}
I(r)= \frac{1}{2\pi}\int_{0}^{2\pi}  i(r,\theta) d \theta,
\end{equation}
where  $i(r,\theta)$ is the local light intensity contained in a circle of radius $r$.  Each value of $I(r)$ represents the sum of the pixel intensities around a circle with radius $r$.

\subsection*{Texture analysis based on gray level co-occurrence matrix (GLCM) }
A gray level co-occurrence matrix (GLCM) is a matrix where the number of rows and columns is equal to the number of gray levels $N_g$ in an image. Its analysis is based on the correlation among pixels in an image. Mathematically, this information is captured by the matrix element $p(i, j)$, which represent the probability values for changes between gray level $i$ and $j$ at a particular displacement distance ($d$) and angle ($\phi$) on an image.  This probability can be defined as:
\begin{equation}
p(i,j) = \frac{C(i,j)}{\sum_{i=0}^{N_g-1} \sum_{j=0}^{N_g-1} C(i,j)},
\end{equation}
where $C(i, j)$  is the number of occurrences of gray levels $i$ and $j$ within the window, at a particular ($d, \phi$) pair. The denominator is the total number of gray level pairs $(i, j)$ within the window and is bounded by an upper limit of $N_g x N_g$. The mean and the standard deviation for the columns and rows  of the matrix, using the above equation,  can be defined as follows:
\begin{equation}
u_x = \sum_{i=0}^{N_g-1} \sum_{j=0}^{N_g-1} i \cdot p(i,j), \hspace{0.5cm} u_y = \sum_{i=0}^{N_g-1} \sum_{j=0}^{N_g-1} j \cdot p(i,j),
\end{equation}
\begin{equation}
\sigma_x= \sum_{i=0}^{N_g-1} \sum_{j=0}^{N_g-1} (i-u_x)^{2} \cdot  p(i,j),  \hspace{0.5cm} \sigma_y= \sum_{i=0}^{N_g-1} \sum_{j=0}^{N_g-1} (j-u_y)^{2} \cdot  p(i,j),
\end{equation}
where $u_x$ and $u_y$ are the mean for the columns and rows, respectively; and $\sigma_x$ and $\sigma_y$ represent the standard deviation for the columns and rows, respectively. Now, using these equations, we can define  the configurational entropy parameter of GLCM as follows: 

\begin{equation}
H = -\sum_{i=0}^{N_g-1} \sum_{j=0}^{N_g-1} p(i,j) log (p(i,j)) .
\end{equation}
Entropy is a  measure of homogeneity of a surface and captures the randomness of the image texture. Higher (lower) entropy values indicate large (small) heterogeneous regions in an image. The more significant the entropy, the more pixels in high contrast.

\subsubsection*{Wetting and drying paper processes}
The paper wetting process was carried out by placing plate of watercolor paper vertically inside a container (10x50x170mm) with pure water for three minutes. To estimate the loss of mass by evaporation, we put the paper on a glass plate (100x100x1mm) placed on a micro analytical balance. The initial relative humidity was controled using silica gel (Sigma-Aldrich, 13767).

The drying paper process evolves in three different stages of evaporation that generate a specific type of water distribution on the paper: water bodies (I), stagnant water (II) and water in bulk between pores and fibers (III), respectively. Fig. 7a shows a sequence of images during the drying of watercolor paper at T=23 $^{\circ}$C. In the first stage (I), the liquid on the surface of the paper forms water puddle that are initially asymmetrical. In regions where water bodies do not form, water stagnates within the cavities formed by the roughness of the paper. As the highest rate of evaporation of water molecules occurs in the periphery of the paper, the bodies of water coalesce until forming a single body of water that evolves until reaching spherical symmetry. In addition, stagnant water emerges on the periphery of the paper advancing toward the center. In the second stage (II), the water body has disappeared and on the surface of the paper dominates the region with stagnant water. Temporarily, this second stage is very small and allows high wetting on the surface of the paper.  Finally, in the last stage of drying (III) there is no stagnant water on the surface of the paper but a large amount of water in bulk between the pores and fibers.

Configurational entropy captures changes in surface water distribution during paper drying, see Fig. 7b. Initially entropy increases due to the evolution and coexistence between puddle of water and stagnant water. The greatest contrast in the texture of wet paper (the maximum configurational entropy value) is observed during the formation of a body of water surrounded by a large region of stagnant water. Clearly, this contrast in texture decreases with the reduction of the radius of the water body. Once the water body disappears, the configurational entropy rapidly decreases due to the very short lifespan of the stagnant water. Finally, the texture of the paper changes slowly during the final drying stage.

Fig. 7c shows the evolution of the surface density of water as a function of time during the drying process of 4 wet papers. The slope of the straight lines that best fit the experimental data provides the evaporation rate, which on average is $-1.405gr/s$ $\pm$0.027. Fig. 7d shows the transition between the three stages of water distribution during the evaporation of water in watercolor paper. Interestingly, we observe that the different water distributions (water bodies (I), stagnant water (II) and water in bulk between pores and fibers (III)) can have the same {surface water density} value. This means that the density of surface water is not an efficient parameter to characterize the relative humidity inside the paper.

\begin{thebibliography}{10}


\bibitem{Van} Van der Weerd, J., Brammer, H., Boon, J. J., \&  Heeren, R. M. A.  Fourier transform infrared microscopic imaging of an embedded paint cross-section. \textit{Appl Spectrosc.} \textbf{56} , 275-283 (2002).

\bibitem{Klockenkamper} Klockenkamper, R., Von Bohlen, A., \&  Moens, L. . Analysis of pigments and inks on oil paintings and historical manuscripts using total reflection x‐ray fluorescence spectrometry. \textit{X-Ray Spectrom.} \textbf{29}, 119-129 (2000).

\bibitem{Burgio} Burgio, L., Clark, R. J., Stratoudaki, T., Doulgeridis, M., \&  Anglos, D.  Pigment identification in painted artworks: a dual analytical approach employing laser-induced breakdown spectroscopy and Raman microscopy. \textit{Appl. Spectrosc.} \textbf{54}, 463-469 (2000).

\bibitem{Brett}  Brett, B.  A history of watercolor. New York: Excalibur Books. (1984).

\bibitem{Mayer}   Mayer, R.  The artist's handbook of materials and techniques. (1991).

\bibitem{Abdulsalam}  Abdulsalam, S., \& Maiwada, Z. D. Production of Emulsion House Paint Using Polyvinyl Acetate and Gum Arabic as Binder. \textit{IJMSA} \textbf{4}, 350 (2015). 


\bibitem{Rounsley} Rounsley, R. R. Vapor transport through paper. \textit{Tappi} \textbf{47}, 95-98 (1964).

\bibitem{Liang} Liang, B., Fields, R.J. \& King, J.C., The Mechanisms of Transport of Water and  n-Propanol through Pulp and Paper, \textit{Drying Tech.} \textbf{8(4)}, 641-665 (1990).

\bibitem{Nilsson1} Nilsson, L., Wilhelmsson, B. \& Stenstrom, S., The Diffusion of Water Vapor  Through Pulp and Paper, \textit{Drying Tech.} \textbf{11(6)}, 1205-1225 (1993).

\bibitem{Nilsson2} Nilsson, L. \& S. Stenstrom. Gas diffusion through sheets of fibrous porous  media. \textit{Chem. Eng. Sci.}  \textbf{50}, 361-374 (1995).

\bibitem{Hashemi} Hashemi, S. J., V. G. Gomes, R. H. Crotogino \& W. J. M. Douglas. In-Plane  diffusivity of moisture in paper. \textit{Drying Tech.} \textbf{15}, 265-294 (1997).



\bibitem{Radhakrishnan}  Radhakrishnan, H., Chatterjee, S. G. \& Ramarao, B. V. Steady-State Moisture  Transport in a Bleached Kraft Paperboard Stack. \textit{J. Pulp Pap. Sci.} \textbf{26},  140-144  (2000). 

\bibitem{Massoquete}  Massoquete, A., S. Lavrykov, B. V. Ramarao \& S. Ramaswamy. The effect of  refining on moisture diffusion in paper. \textit{ESPRA Res. Report} \textbf{118},  64-87 (2003).




\bibitem{Harting2016} Liu, H., Kang, Q., Leonardi, C. R., Schmieschek, S., Narv{\'a}ez, A., Jones, B. D., Williams, J. R., Valocchi, A. J., \& Harting, J. Multiphase lattice Boltzmann simulations for porous media applications. \textit{Computational Geosciences}. \textbf{20}, 4:777--805, (2016).


\bibitem{Nilsson3}  Nilsson, L.  Mansson, S. \&  Stenstrom, S; Measuring Moisture Gradients in  Cellulose Fiber Networks an Application of the Magnetic-Resonance-Imaging  Method \textit{	J. Pulp Pap. Sci.} \textbf{22}, J48-52 (1996).




\bibitem{Li}  Li, T. Q., Haggkvist M., \&  Odberg L. The porous structure of paper coatings studied by water diffusion measurements. \textit{Colloids Surf., A} \textbf{159},  57-63 (1999).

\bibitem{Harding}  Harding, S. G., Wessman, D., Stenstrom, S., \& Kenne, L.  Water transport during the drying of cardboard studied by NMR imaging and diffusion techniques. \textit{Chem. Eng. Sci.} \textbf{56}, 5269-5281 (2001).

\bibitem{Scallan}  Scallan, A. M.  On no solvent water in cellulosic fibers as determined by salt exclusion. \textit{Cellul. Chem. Technol.} \textbf{21}, 215-223 (1987).


\bibitem{Maloney}   Maloney, T. C., \& Paulapuro, H. The formation of pores in the cell wall. \textit{JPPS} \textbf{25}, 430-436 (1999). 


 

\bibitem{deegan2000contact}  Deegan, R. D. Bakajin, O., Dupont, T.F., Huber, G., Nagel, S.R. \& Witten, T.A., Contact line deposits in an evaporating drop. \textit{Phys. review E} \textbf{62}, 756 (2000).



\bibitem{hu2006marangoni}     Hu, H. \& Larson, R. G. Marangoni effect reverses coffee-ring depositions. \textit{J. Phys. Chem. B} \textbf{110}, 7090-7094 (2006).







\bibitem{deegan1997capillary} Deegan, R. D. \textit{et al.} Capillary flow as the cause of ring stains from dried liquid drops. \textit{Nature} \textbf{389}, 827-829 (1997).



\bibitem{Marin2011:order} Marin, A., Gelderblom, H., Lohse, D., Snoeijer, J. H. Order-to-Disorder Transition in Ring-Shaped Colloidal Stains. \textit{Physical Review Letters}  107, 085502 (2011).


\bibitem{still2012surfactant}  Still, T., Yunker, P. J. \& Yodh, A. G. Surfactant-induced marangoni eddies alter the coffee-rings of evaporating colloidal drops. \textit{Langmuir} \textbf{28}, 4984-4988 (2012).

\bibitem{sempels2013auto}  Sempels, W., De Dier, R., Mizuno, H., Hofkens, J. \& Vermant, J. Auto-production of biosurfactants reverses the coffee ring effect in a bacterial system. \textit{Nat Commun.} \textbf{4}, 1757 (2013).


\bibitem{Marin2016:surfactant} Marin, A., Liepelt, R., Rossi, M., K\"ahler, C. J. Surfactant-driven Flow transitions in evaporating droplets. \emph{Soft Matter} 12, 1593 (2016).



\bibitem{Marin2019:salt} Marin A. 
Karpitschka, S., Noguera-Mar\'in, D., Cabrerizo-V\'ilchez, M. A., Rossi, M., K\"ahler, C. J., Rodr\'iguez Valverde, M. A. Solutal Marangoni flow as the cause of ring stains from drying salty colloidal drops. \textit{Physical Review Fluids} 4, 041601(R) (2019).



\bibitem{hu2005marangoni}
Hu, H. \& Larson, R. G. Analysis of the Effects of Marangoni Stresses on the Microflow in an Evaporating Sessile Droplet. \textit{Langmuir} 21, 3972–3980 (2005).













\bibitem{gorr2014pattern}  Gorr, H. M., Xiong, Z. \& Barnard, J. A. Pattern recognition for identification of lysozyme droplet solution chemistry. \textit{Colloids Surf., B} \textbf{115}, 170-175 (2014).


\bibitem{chen2010complex} Chen, G. \& Mohamed, G. J. Complex protein patterns formation via salt-induced self-assembly and droplet evaporation. \textit{The Eur. Phys. J. E} \textbf{33}, 19-26 (2010).

\bibitem{gorr2012characteristic} Gorr, H. M., Zueger, J. M. \& Barnard, J. A. Characteristic size for onset of coffee-ring effect in evaporating lysozyme-water solution droplets. \textit{ J. Phys. Chem. B} \textbf{116}, 12213-12220 (2012).

\bibitem{yakhno2011complex} Yakhno, T. A. Complex pattern formation in sessile droplets of protein-salt solutions with low protein content. what substance fabricates these patterns? \textit{ Phys. Chem.} \textbf{1}, 10-13 (2011).


\bibitem{carreon2017patterns}  Carre\'on, Y. J., Gonz\'alez-Guti\'errez, J., P\'erez-Camacho, M. \& Mercado-Uribe, H. Patterns produced by dried droplets of protein binary mixtures suspended in water. \textit{Colloids Surf., B} \textbf{161}, 103-110 (2018).

\bibitem{pauchard1999influence}   Pauchard, L., Parisse, F. \& Allain, C. Influence of salt content on crack patterns formed through colloidal suspension desiccation.\textit{ Phys. Rev. E} \textbf{ 59}, 3737 (1999).

\bibitem{Sefiane} Sefiane, K. "On the formation of regular patterns from drying droplets and their potential use for bio-medical applications." \textit{JBE} \textbf{7} S82-S93  (2010).

\bibitem{Nellimoottil}  Nellimoottil, T. T., Rao, P. N., Ghosh, S. S., \& Chattopadhyay, A.  Evaporation-induced patterns from droplets containing motile and nonmotile bacteria.\textit{ Langmuir}, \textbf{23}, 8655-8658 (2007).

\bibitem{Rios}   R\'ios-Ram\'irez, Maricarmen, M., Reyes-Figueroa, A.D., Ruiz-Su\'arez,\&  J.C. and Gonz\'alez-Guti\'errez, Pattern formation of stains from dried drops to identify spermatozoa motility. \textit{Colloids Surf., B} \textbf{169} 486-493 (2018).

\bibitem{gonzalez2017calorimetric}  Gonz\'alez-Guti\'errez, J., P\'erez-Isidoro, R., P\'erez-Camacho, M. I. and Ruiz-Su\'arez, J. The calorimetric properties of liposomes determine the morphology of dried droplets. \textit{Colloids Surf., B} \textbf{155}, 215-222 (2017).

\bibitem{gonzalez2017technique} Gonzalez-Guti\'errez, J., P\'erez-Isidoro, \& R. \& Ruiz-Su\'arez, J. A technique based on droplet evaporation to recognize alcoholic drinks. \textit{Rev. Sci. Instruments} \textbf{88}, 074101 (2017).


\bibitem{gonzalezproteins} Carre\'on, Y. J., R\'ios-Ramírez, M., Moctezuma, R. E., \& Gonz\'alez-Guti\'errez, J.  Texture analysis of protein deposits produced by droplet evaporation. \textit{Scientific Reports}, \textbf{8}, 9580. (2018).

\bibitem{Marin2011:rush} Marin, A., Gelderblom, H., Lohse, D., Snoeijer, J. H. 
Rush-hour in evaporating coffee drops. \emph{Physics of Fluids} 23, 091111 (2011)







\bibitem{Gelderblom2012} Gelderblom, H., Bloemen, O., \& Snoeijer, J. H.. Stokes flow near the contact line of an evaporating drop. \emph{Journal of Fluid Mechanics} 709: 69-84(2012).




\end{thebibliography}

{}

\section*{Acknowledgements}

J.G.G. wish to acknowledge financial support by DGAPA-UNAM. We  acknowledge  C. Ruiz-Suarez for  the use of the z-sizer.  The support of DGAPA-PAPIIT-UNAM (grant number IN108016) and ACT-FONCA (grant number 04S.04.IN.ACT.038.18) are greatly acknowledged.

\section*{Author contributions statement}

J.G.G. conducted the experiments, analysed the data, and elaborated all the Figures. J.G.G. and V. A. analysed the evaporation data.   M.R.R conducted particle size measurements.  J.G.G., A.M. and R.Z. discussed results.  J.G.G., A. M. and R.Z. wrote the paper.  All authors reviewed the manuscript. 

\section*{Additional information}

\textbf{Competing Interests: }The authors declare no competing interests.


\begin{figure*}[!ht]
\begin{center}
\includegraphics[width=\linewidth]{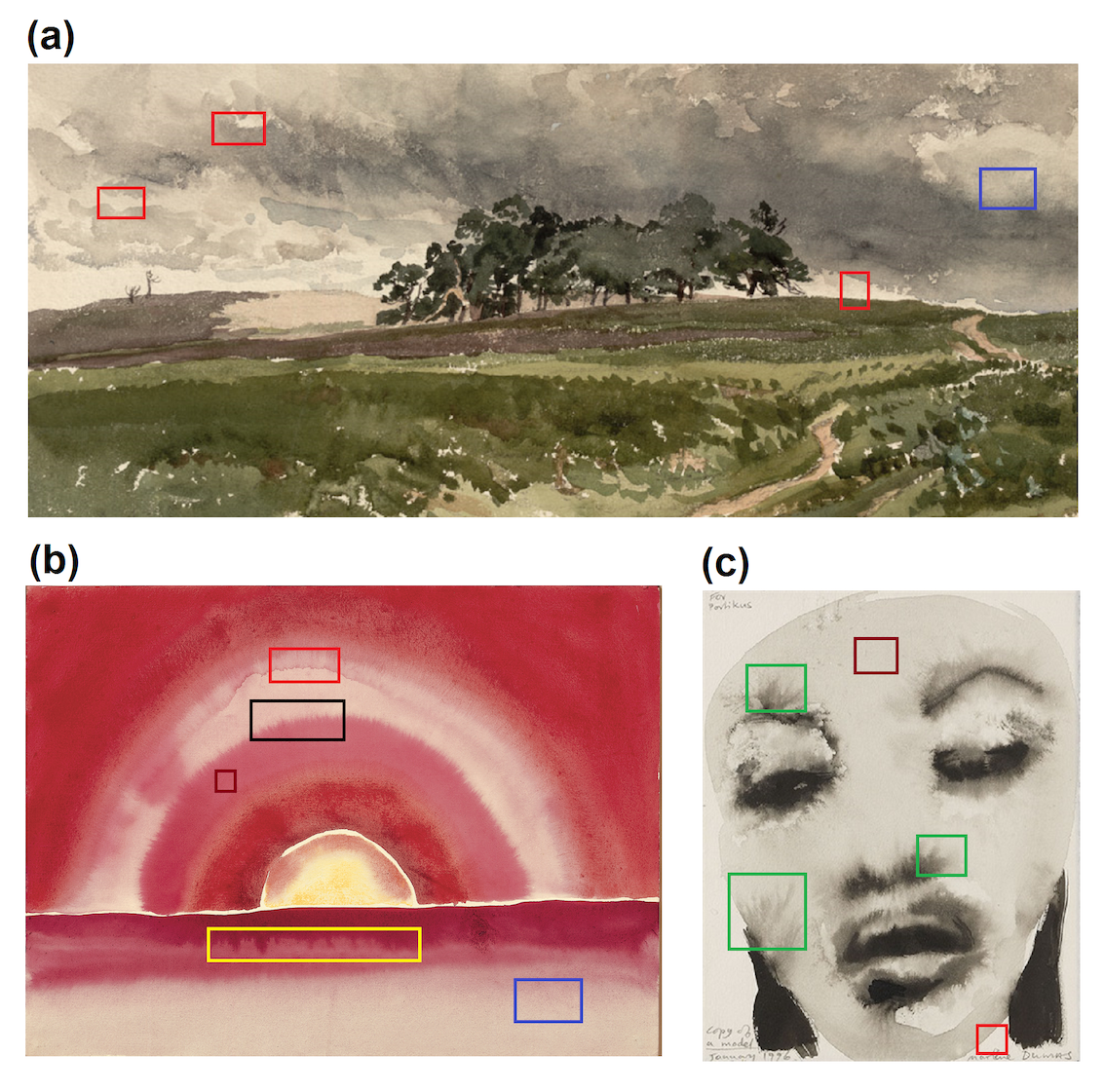}
\end{center}
\caption {\textbf{Watercolor Paintings} Examples of water colour paintings,  from classical and realist to modern and abstract. (a) \textit{A Heath scene} by T. Collier, 1860, (b) \textit{Sunrise} by G. O' Keeffe, 1916; and (c) \textit{Model} by M. Dumas, 1996. In the paintings we find different characteristic patterns (depicted by the coloured boxes): sharp edges (red), gradients of color (blue), flat surfaces (wine), small radial tips (black), large radial tips (green), and dendrites (yellow).}
\label{gr1}
\end{figure*}
\begin{figure*}[!ht]
\begin{center}
\includegraphics[width=\linewidth]{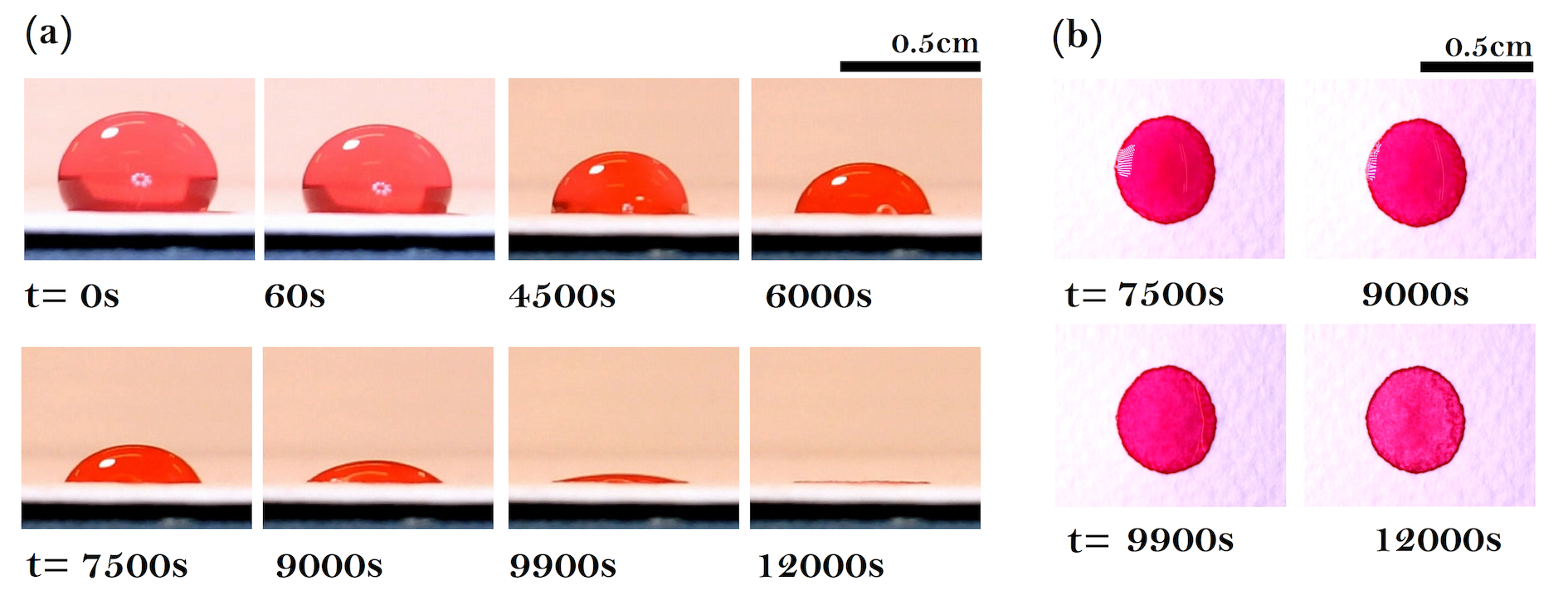}
\end{center}
\caption {\textbf{The pattern formation by drop evaporation on dry paper.} (a) Top and (b) side view of the pattern formation from the  evaporation of a droplet of ``genuine pink'' ($\phi_p$ = 1 wt\% , $\theta$ = 128 $^\circ$,  and T = 22 $^{\circ}$C). }
\label{gr2}
\end{figure*}



\begin{figure*}[!ht]
\begin{center}
\includegraphics[width=\linewidth]{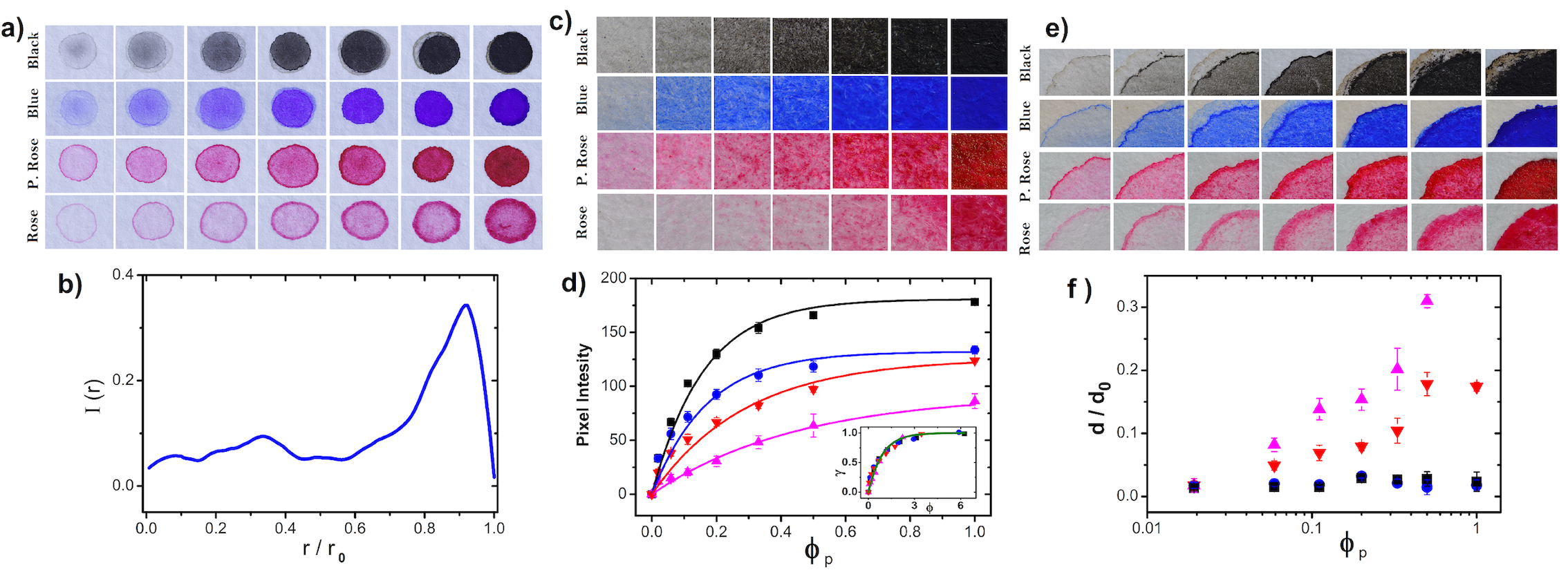}
\end{center}
\caption {\textbf{The effect of the pigment concentration on the pattern formation by drop evaporation on dry paper.} (a) Deposits formed after the evaporation of droplets of four different types of pigments at concentrations $\phi_p$ = 0, 0.05, 0.1, 0.25, 0.5, and 1 wt\% and T = 22 $^{\circ}$ C. (b)  The radial density profile as a function of the relative radius (where $r_o$ is the maximum radius of the stain) of a deposit pattern formed by permanent ``pink pigment'' at  $\phi_p$=0.5.  (c) The central regions    of the patterns in (a). (d) The mean pixel intensity  as a function of the relative  pigment concentration. (e) The coffee ring thickness  of the patterns in (a). (f) The coffee ring thickness  as a function of the relative pigment concentration. The Square, circle, triangle, and inverted triangle indicate black, blue, Permanent rose and rose genuine, respectively. }
\label{gr3}
\end{figure*}

\begin{figure}[!ht]
\centering
\includegraphics[width=\linewidth]{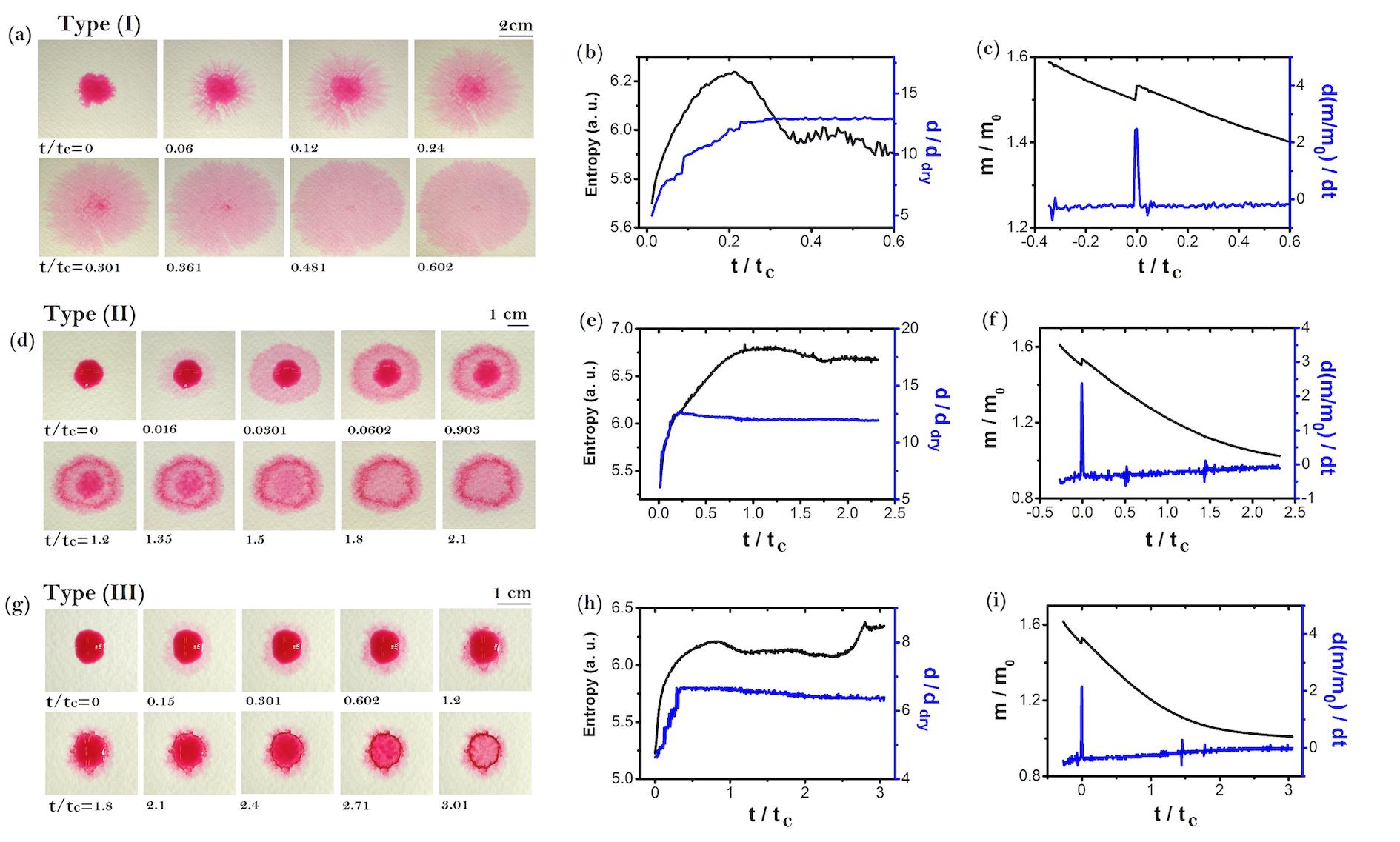}
\caption{ \textbf{Pattern formation of watercolor ink containing colloidal particles.}  (a) Image sequence of the patter growth in the type (I)  ($\phi_p$ = 1 wt\% and T = 22 $^{\circ}$C). (b) The corresponding evolution of the entropy and the diameter. (c) The superficial water density as a function of the time. (d) The patter formation in the type (II) ($\phi_p$ = 1 wt\% and T = 22 $^{\circ}$C). (e) The entropy and the diameter evolution during the pattern formation . (f) The superficial water density as a function of the time. (g) A sequence of images of the patter formation in the type (III) ($\phi_p$ = 1 wt\% and T = 22 $^{\circ}$C). (h) The evolution of the entropy and the diameter (blue line). (i) The relative mass $m/m_0$ as a function of $t/t_c$. } 
\end{figure}

\begin{figure}[!ht]
\centering
\includegraphics[width=\linewidth]{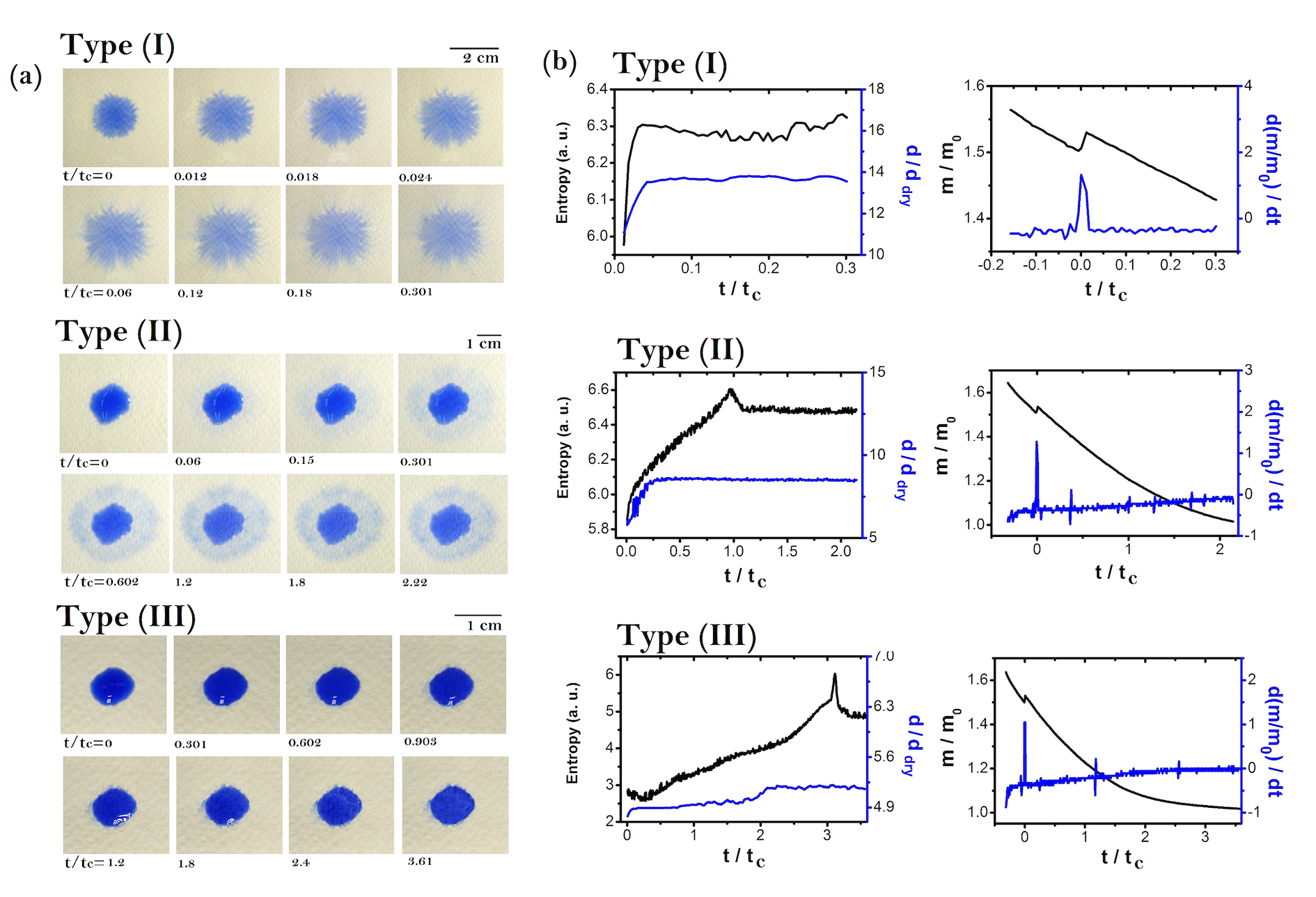}
\caption{ \textbf{Pattern formation of watercolor ink containing non-colloidal particles.}   (a) The pattern formation from the evaporation of watercolor droplets  at different  type of water distribution ($\phi_p$ = 1 wt\% and T = 22 $^{\circ}$C). (b) The corresponding evolution of entropy, diameter, and relative mass $m/m_0$ of (a). } 
\end{figure}

\begin{figure}[!ht]
\centering
\includegraphics[width=\linewidth]{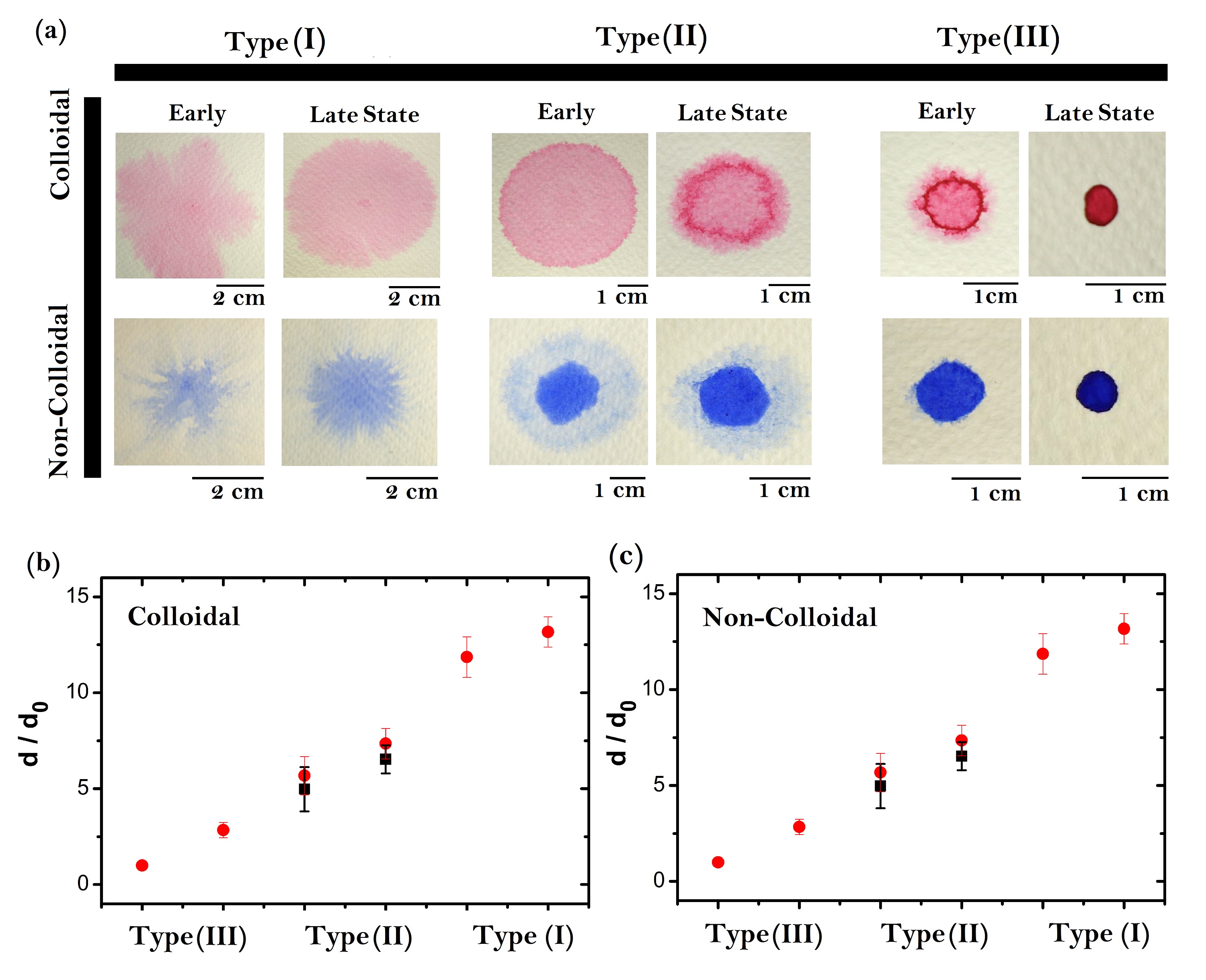}
\caption{ \textbf{Watercolor patterns formed from the drop deposition in the early and late stage of drying of the three water distributions.}  (a) Patterns  formed at the same water density (0.0306 $g/cm^2$), $\phi_p$ = 1 wt\%  and T= 25 $^\circ$C but different types of water distribution.  The labels ``early'' and ``late'' indicate that the stains are produced in early and late evaporation stages of paper, respectively, but with the same type of water distribution and content of water in paper.  (b) and (c) diameter of stains for the different types of water distribution for watercolor inks containing Brownian and non-Brownian pigments, respectively. } 
\end{figure}

\begin{figure}[!ht]
\centering
\includegraphics[width=\linewidth]{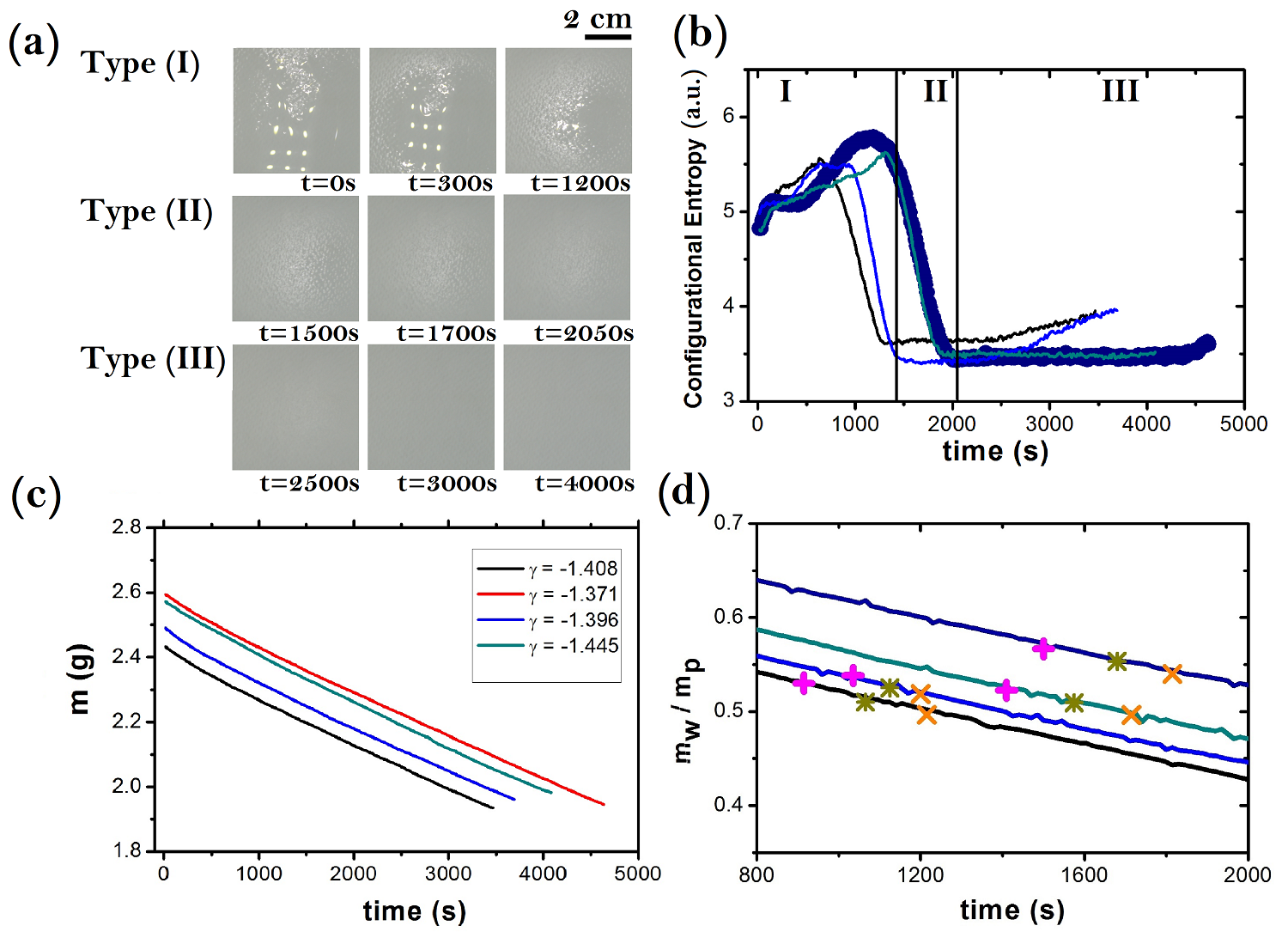}
\caption{\textbf{Paper drying process.} (a) The three types of water distribution on the paper that  emerge  in the three stage of evaporation. (b) The Entropy evolution during the drying process. (c) The mass as a function of the time. (d) A zoom of the relative mass of water as a function of the time (where $m_w$ and $m_p$ are the mass of water and paper, respectively). The cross, asterisk and cross out corresponds to evaporation stage I, II, and III, respectively.} 
\end{figure}
\end{document}